\documentclass[twocolumn,showpacs,prl]{revtex4}

\usepackage{graphicx}
\usepackage{dcolumn}
\usepackage{bm}

\begin{document}

\setlength{\baselineskip}{0.4cm} \addtolength{\topmargin}{1.5cm}

\title{Avalanche prediction in Self-organized systems}

\author{O. Ramos$^{1,2}$}
\email{osvanny.ramos.rosales@ens-lyon.fr}
\author{E. Altshuler$^{3}$}
\author{K. J. M{\aa}l{\o}y$^{1}$}

\affiliation{ $^1$Department of Physics, University
of Oslo, P.O.B. 1048, Blindern N-0316, Oslo, Norway\\
$^2$Laboratoire de Physique, CNRS UMR 5672, Ecole Normale
Sup\'{e}rieure de Lyon, 46 all\'{e}e d'Italie, 69364 Lyon Cedex 07,
France\\
$^3$``Henri Poincar\'e" Group of Complex Systems, Physics Faculty,
University of Havana, 10400 Havana, Cuba }

\date{\today}

\begin{abstract}

It is a common belief that power-law distributed avalanches are
inherently unpredictable. This idea affects phenomena as diverse as
evolution, earthquakes, superconducting vortices, stock markets,
etc; from atomic to social scales. It mainly comes from the concept
of ``Self-organized criticality" (SOC), where criticality is
interpreted in the way that at any moment, any small avalanche can
eventually cascade into a large event. Nevertheless, this work
demonstrates experimentally the possibility of avalanche prediction
in the classical paradigm of SOC: a sandpile. By knowing the
position of every grain in a two-dimensional pile, avalanches of
moving grains follow a distinct power-law distribution. Large
avalanches, although uncorrelated, are preceded by continuous,
detectable variations in the internal structure of the pile that are
monitored in order to achieve prediction.

\end{abstract}


\maketitle

In the last two decades many efforts have been devoted to understand
the ubiquity of scale invariance in nature. The best known attempt
so far, although controversial, has been the Self-organized
criticality (SOC) \cite{BTW 1987, Jensen 1998} where the competition
between a driving force that very slowly injects energy and a
dynamics of local thresholds, can drive the system into a critical
state where a minor perturbation can trigger a response (avalanche)
of any size and duration. A major goal of these kinds of models is
to gain understanding of processes ruled by scale invariance that
eventually have catastrophic events (evolution \cite{Sneppen et al
1995}, earthquakes \cite{OFC 1992, Ramos et al 2006}, stock markets
\cite{Lee et al 1998}, solar flares \cite{Hamon et al 2002},
superconducting vortices \cite{Altshuler & Johansen 2004, Altshuler
et al 2004}, etc), in order to predict them. Since unpredictability
is an essential feature of critical processes evolving through
power-law distributed avalanches, the possibility of prediction
seems unlikely. However, prediction is -in principle- possible,
because the system is not at, but just close to, the critical state
\cite{Carvalho and Prado 2000, Comment}. Some cellular automaton
models of earthquakes have analyzed the predictability of very large
avalanches (responsible for the cut-off on a power-law distribution)
\cite{Pepke and Carlson 1994} and also precursors of large events
have been reported in dissipative or hierarchical lattices
\cite{Huang et al 1998, Sammis & Smith 1999}. However, no
experimental results have been presented so far concerning
prediction of power-law distributed avalanches in slowly driven
systems; and still many researchers think that if earthquakes
correspond to a SOC process, they are inherently unpredictable
\cite{Geller et al 1997, Main 1999}.

\begin{figure}[b!]
\vspace{-0.4cm}
\includegraphics[height=2.3in, width=3.4in]{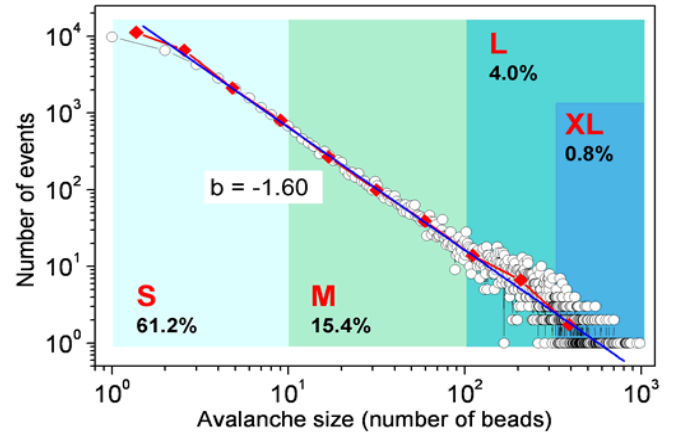}
\vspace{-0.3cm} \caption{\label{ava-dist} (color online)
Distribution of avalanche size (open circles). These points have
been averaged with a logarithmic binning (diamonds). Avalanches are
classified in small (S), medium (M), large (L) and very large (XL)
considering logarithmic bins. The percentage of every type of
avalanche related to the total number of dropping events is also
displayed.}
\end{figure}

\begin{figure}[t!]
\includegraphics[height=3.3in, width=3.4in]{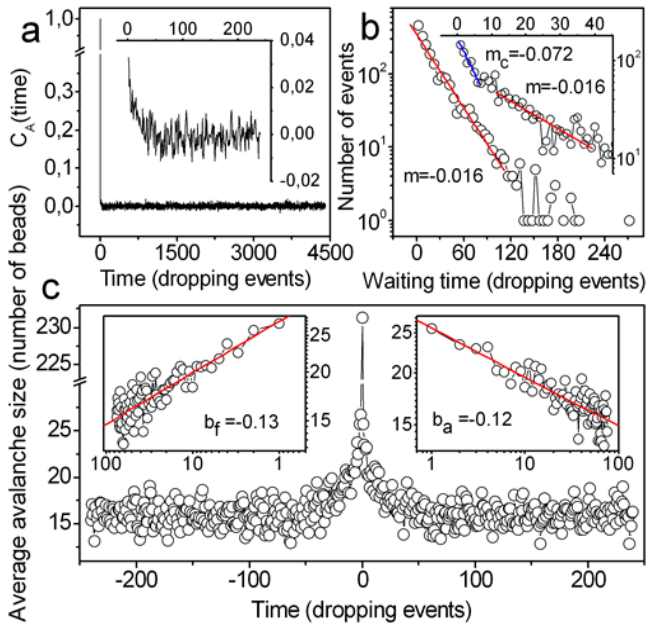} 
 \caption{\label{time-series} (color online) Analysis of the avalanche time series.
  (a) Autocorrelation of the large avalanche time series. (b)
Distributions of waiting time between large avalanches. (c) Average
avalanche size around a large avalanche.}
\end{figure}

We present an experiment where avalanches display a power-law
distribution with an exponent equal to -1.6 and where precursors to
large events have been found. We use a setup similar to that
reported by Altshuler et al. \cite{Altshuler et al 2001} where SOC
behavior was observed. However, the present experiment is the first
one having single grain resolution in the measurement of the
avalanches that take place not only at the free surface \cite{Frette
et al 1996, Aegerter et al 2003} or falling off the system
\cite{Altshuler et al 2001, Held et al 1990, Rosendahl et al 1994,
Costello et al 2003}, but also in the bulk of the pile. A base,
consisting of a 60 cm long row of 4 $\pm$ 0.005 mm steel spheres
separated from each other by random (0, 1, 2 or 3 mm) spacing, is
glued to an acrylic surface and sandwiched between two parallel
vertical glass plates 4.5 mm apart. The same steel beads are
delivered one by one from a height of 28 cm above the base and at
its center, resulting in the formation of a quasi-two-dimensional
pile. The extremes of the base are open, leaving the beads free to
abandon the pile. After a bead is delivered, the system waits for a
few seconds in order to guarantee that all the relaxation effects in
the pile are finished. The pile is then recorded with a Canon D20
digital camera at a resolution of 21 pixels/bead diameter, followed
by the dropping of a new bead. One experiment contains more than
55000 dropping events with a total duration of more than 310 hours.
The first 4500 events before the pile reaches a stationary state are
not included in the statistics. The average number of beads in the
pile is 3315. By processing the images, the centers of all the beads
are found and the avalanche size is defined as the number of beads
moving between two consecutive dropping events (we include here the
beads that fall off the pile). We have assumed that one bead has
moved when its center does not have any neighboring center at a
distance less than or equal to 1/7 of the bead diameter in the
consecutive image. The distribution of avalanche sizes is displayed
in Fig.~\ref{ava-dist}. Avalanches span over three decades in a
log-log plot and follow a power-law in almost all this range. The
graph can be divided (without losing generality) in three equally
spaced zones. So, in considering a distribution that spans over $3n$
decades, avalanches smaller than $n$ are considered small, those
lying between $n$ and $2n$ are medium, and those greater than $2n$
are large. We focus our attention on the large ones (size$>$100) and
we look for precursors in the structure of the system. The very
large avalanches (size$>$316), which have a very small probability
of occurrence, are also analyzed.

\begin{figure}[b!]
\vspace{-0.4cm}
\includegraphics[height=4.09in, width=3.3in]{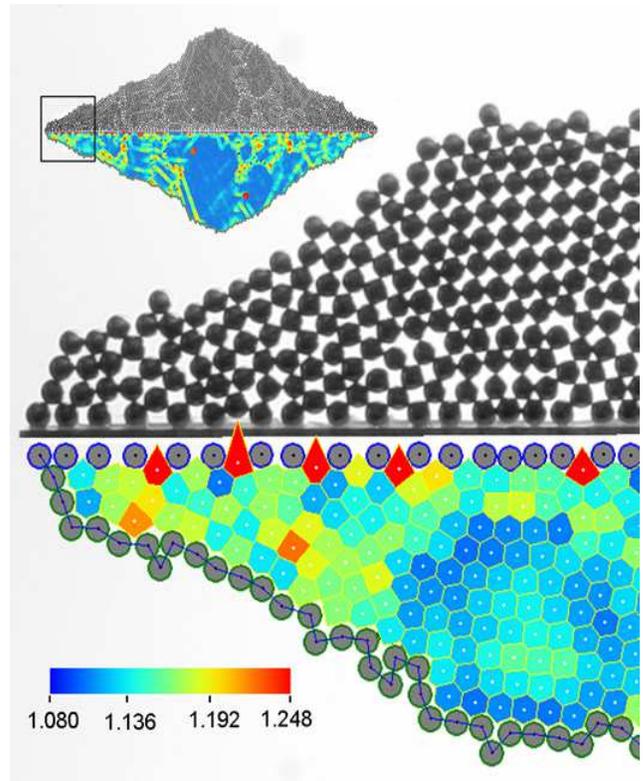}
\caption{\label{internal-struct}(color online) Internal structure of
a pile. Close up of a portion of the pile with a reflection (below)
in the space of the shape factor $\zeta$. After finding the center
of the beads and making a voronoi triangulation for the internal
ones (i.e. excluding those in the base and those joined with a line
at the pile's surface) we define  $\zeta=C^{2}/4\pi S$, where $C$ is
the perimeter and $S$ the area of each voronoi cell. The shape
factor $\zeta$ is a measure of the local disorder in the pile (for
example for a regular hexagon $\zeta$=1.103 (highest order in the
pile) and for a square $\zeta$=1.273). Inset: the whole pile, where
a rectangle indicates the closed up area. Notice the small values of
$\zeta$ at the center of the pile, indicating hexagonal packing
which results from the ``thermal" effect of beads being added from
above.}
\end{figure}

\begin{figure*}[t!]
\includegraphics[height=3.193in, width=6.613in]{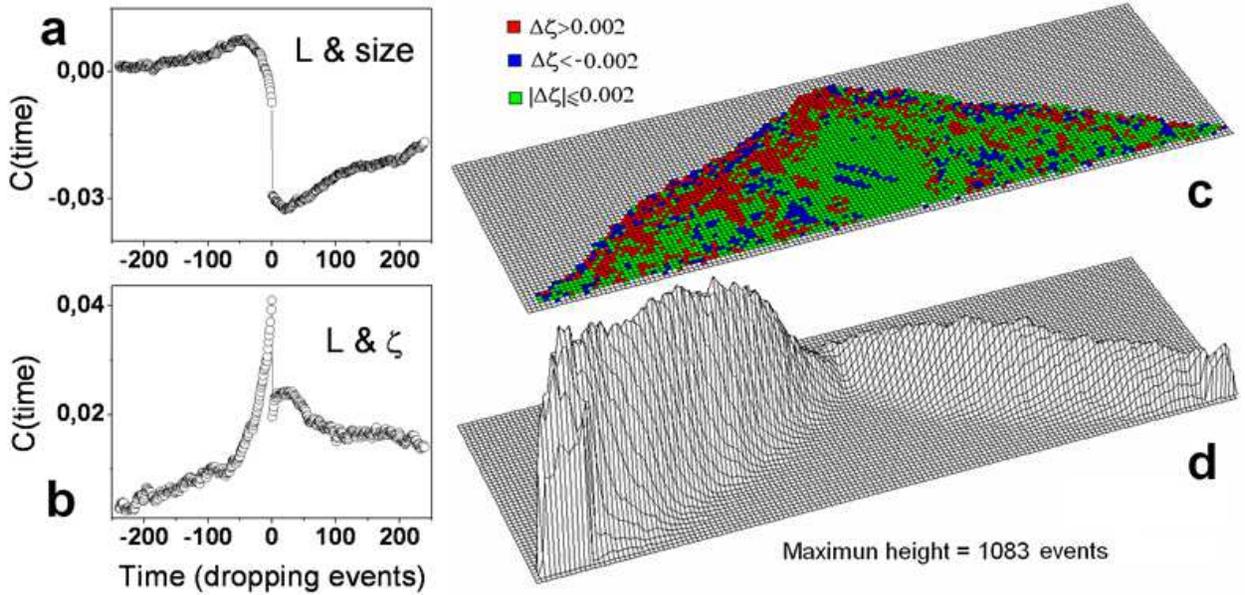}
\caption{\label{correlations} Correlation between avalanches $\&$
internal structure. (a) Temporal correlation between the large
avalanches and the size of the system. (b) Temporal correlation
between the large avalanches and the average shape factor $\zeta$.
(c) Difference between the local averages of the $\zeta$ values at
one step before a large avalanche and at fifty steps before a large
avalanche. Red indicates that the disorder increases, blue an
increase in the order, and green displays no variations in $\zeta$.
A cursory inspection shows that the red color predominates over the
blue. The cumulative number of sites involved in large avalanches is
displayed at (d). The match between the red color in (c), and the
landscape in (d), corroborates that on average, the pile suffers an
increment of the disorder before a large avalanche takes place.}
\end{figure*}

The temporal autocorrelation of large avalanches reads as
\begin{equation}
  \label{eq2}
  C_A(t)=\frac{\sum ( s(\tau) s(\tau+t))-<s(\tau)>^{2}}{\sum ( s(\tau)-<s(\tau)>)^{2}}
\end{equation}
\noindent where $s(\tau)$ equals unity if the avalanche is large,
and zero in all other cases. The unit of time corresponds to one
dropping event, also called step. The uncorrelated character of
large (L) avalanches is displayed in Fig.~\ref{time-series}$a$. The
waiting time between them follows an exponential distribution
(Fig.~\ref{time-series}$b$). This implies the presence of a
characteristic waiting time, equal to 26 $\pm$ 2 steps, indicating
the average time between large events. Very large avalanches (XL)
behave in a similar way, with a characteristic waiting time equal to
132 $\pm$ 9 steps. Concerning predictability, it is interesting to
note that the earthquake scenario is better than the one presented
here, in the way that the waiting times between events follow a
power-law distribution, allowing some global, or long term, forecast
\cite{Corral 2004, Corral 2005}. Up to this point in the analysis,
we have a self-organized system with uncorrelated avalanches whose
sizes follow a power-law distribution. So it seems impossible to
predict the moment when a large avalanche is going to happen.
However, a slight decay larger than the background noise can be
noticed for the first 20 steps in the inset of
Fig.~\ref{time-series}$a$. This is an indication of a time
clustering of avalanches, and a steeper slope $m_c$ for the small
values of the waiting time in the inset of Fig.~\ref{time-series}$b$
is also a sign of it.  This temporal clustering is analyzed in
Fig.~\ref{time-series}$c$: around a large event there is an average
increment of the avalanche size. The best fit of it corresponds to
power-laws (insets) for both``foreshocks" and ``aftershocks". The
exponents of the power-laws are very low (-0.13 and -0.12), making
it impossible to predict main shocks by analyzing possible
foreshocks. Any comparison with the Omori law \cite{Omori 1895} for
earthquakes can be done just qualitatively due to the fact that in
our experiment there is only one avalanche per unit time.

To predict, in the short-term, when a large avalanche is going to
happen, we analyze the corresponding changes in the internal
structure of the pile. As the position of the centers of all the
particles at every step of the experiment is known, we are able to
define several structural variables and analyze how they evolve
during time, particularly in the neighborhood of a large avalanche.
In this paper we focus our study on just two: the size of the
system, defined as the number of beads in the pile, and the shape
factor $\zeta$ \cite{Mouca & Nezbeda 2005}, which is a measure of
the local disorder in the system (Fig.~\ref{internal-struct}). It is
expected that during a big avalanche, some of the grains will
eventually abandon the pile, and therefore the size of the system
should, on average, decrease. If so, before a large event, the pile,
and thus the amount of energy ready to be released during the
avalanche has to be large. The temporal correlation function

\begin{equation}
  \label{eq3}
  C(t)=\frac{\sum ( s(\tau) x(\tau+t))-<s(\tau)><x(\tau)>}{\sqrt{\sum
   ( s(\tau)-<s(\tau)>)^{2}\sum ( x(\tau)-<x(\tau)>)^{2}}}
\end{equation}

\noindent where $x(\tau)$ corresponds either to the temporal series
of the size of the pile (Fig.~\ref{correlations}$a$) or to the
average shape factor (Fig.~\ref{correlations}$b$), displays the
behavior of these structural variables in the vicinity of a large
avalanche. The number of beads in the pile
(Fig.~\ref{correlations}$a$) behaves as ``expected" and
approximately fifty dropping events (steps) before a large avalanche
the size of the pile, on average, suffers a slight increment and
then a continuous decrement (foreshocks zone) until the avalanche
takes place. During the avalanche the pile's size jumps down. It
then continues decreasing for around 25 steps (aftershocks zone),
but starts to grow again soon after due to the addition of new
grains from the top. The continuous variation displayed by the
disorder of the pile before a large avalanche is much clearer
(Fig.~\ref{correlations}$b$) and approximately fifty steps before a
large event, the average disorder continuously increases until the
avalanche takes place. Then the pile reorganizes itself, but it gets
trapped in an intermediate level of disorder. In the aftershocks
zone the disorder increases, and after that, the pile slowly evolves
into more organized states. Even locally it can be noticed as a very
good match between the zones involved in large avalanches, and the
increase of disorder before the avalanche takes place
(Fig.~\ref{correlations}$c, d$). The observed asymmetry is a
consequence of a slight tilt of the ordered (hexagonal) block at the
center of this particular pile, favoring the movement of grains to
the left of the system. The same general behavior of structural
changes in the pile preceding large events has been observed in
other piles developed under similar conditions.

In order to try an actual prediction, we define an alarm for an
upcoming large (L) event. The alarm is in its {\it on} state when
the differences of the spatial average of the shape factor between
the current time and 50 steps before the current time is larger than
zero. After 50000 events, the alarm is {\it on}  $51 \pm 3 \%$ of
the total time, and $62 \pm 4\%$ of the large avalanches take place
during this situation. The same analysis for the very large (XL)
avalanches gives  $64 \pm 7 \%$ of the events happening under an
alarm that is {\it on} $51 \pm 3 \%$ of the total time. The errors
have been calculated by dividing the data in 10 different portions
of 5000 events each. The mean value and the standard deviation are
the values reported before. Perhaps this 62\% it is not a very
impressive percent (in a random process 50\% of the events will
occur under an alarm that is {\it on} 50\% of the time), but the
first criterion we have used for turning the alarm {\it on} is very
simple. If we use a bit more sophisticated criterion, better results
are obtained. For example, taking into account the existence of
aftershocks, we can define a second alarm that is turned {\it on}
immediately after a large (L) event and has a duration of twelve
steps. If the system is under alarm (either this second alarm or the
original one) $50.0 \pm 0.1 \%$ of the total time (in order to get
this value the thresholds of differences of shape factors in the
original alarm have to be readjusted), $65 \pm 4 \%$ of large
avalanches take place when the alarm is {\it on}. Our results
clearly show that large avalanches can be predicted in our system.
We also believe that this percentage of success can be increased by
using smarter techniques such as pattern recognition tools.

We have presented a self-organized system that evolves through
uncorrelated avalanches distributed following a power-law, and where
signals of foreshocks and aftershocks are too weak for developing
any forecast of large events. However, the analysis of the internal
structure of the system demonstrates that not any small avalanche
can cascade into a large event, a fact that is in agreement with
Olson and Allen \cite{olson & Allen 2005} . Large avalanches require
certain conditions in the structure of the system that have to be
developed; and by monitoring this developing process large events
can, in principle, be forecasted.

We thank O. Duran, G. L{\o}voll, D. Molenaar, S. Santucci, L. Vanel
and K. P. Zabielski. This work was supported by NFR, the Norwegian
Research Council, through a grant by Petromax and SUP.


\begin{thebibliography}{99}
\bibitem{Comment} As dissipation increases large avalanches become
more rare. The system spends larger time intervals far from
situations where a large event can reach the size of the system.
\end{thebibliography}
\end{document}